\documentclass[aps,prl,twocolumn,groupedaddress,amsmath,amssymb]{revtex4}

\usepackage{graphicx}% Include figure files
\usepackage{dcolumn}% Align table columns on decimal point
\usepackage{bm}% bold math
\usepackage{soul}
\usepackage{color,xcolor}
\soulregister{\cite}7
\begin{document}

\title{Metal deficient AlB$_{2}$-type (Ti$_{0.2}$Zr$_{0.2}$Hf$_{0.2}$Nb$_{0.2}$Ta$_{0.2}$)$_{1-\delta}$B$_{2}$ high-entropy diborides with high hardness}

\date{\today}
\author{Wuzhang Yang$^{1,2,3}$}
\author{Qinqing Zhu$^{1,2,3}$}
\author{Guorui Xiao$^{2,3,4}$}
\author{Zhi Ren$^{2,3}$\footnote{Email: renzhi@westlake.edu.cn}}

\affiliation{$^{1}$Department of Physics, Fudan University, Shanghai 200433, P. R. China}
\affiliation{$^{2}$Department of Physics, School of Science, Westlake University, 18 Shilongshan Road, Hangzhou 310024, P. R. China}
\affiliation{$^{3}$Institute of Natural Sciences, Westlake Institute for Advanced Study, 18 Shilongshan Road, Hangzhou 310024, P. R. China}
\affiliation{$^{4}$School of Physics, Zhejiang University, Hangzhou 310058, P. R. China}

\begin{abstract}
We report the synthesis and characterization of metal deficient (Ti$_{0.2}$Zr$_{0.2}$Hf$_{0.2}$Nb$_{0.2}$Ta$_{0.2}$)$_{1-\delta}$B$_{2}$ high-entropy diborides (HEBs).
A single homogeneous AlB$_{2}$-type phase is successfully obtained over the $\delta$ range of 0.03 $\leq$ $\delta$ $\leq$ 0.18.
With increasing $\delta$, the unit-cell volume exhibits a nonmonotonic variation with a maximum at $\delta$ = 0.07.
These metal-deficient HEBs possess high Vickers hardness of 16.6-18.9 GPa at a load of 9.8 N and their phase stability is attributed to the increased mixing entropy.
Our results not only present the first series of metal-deficient AlB$_{2}$-type HEBs, but also suggest the existence of similar multicomponent diborides.\\
\end{abstract}

\maketitle
\maketitle

\section{1. Introduction}
AlB$_{2}$-type high-entropy diborides (HEBs) have attracted a lot of attention as the first high-entropy non-oxide ceramics \cite{HEB1}.
This family of HEBs have a layered hexagonal structure, which consists of two-dimensional transition metal and boron layers stacked alternatively along the $c$-axis.
In each transition metal layer, five or more equimolar group IIIB, IVB and VB elements are randomly distributed at the same crystallographic site, leading to a large mixing entropy.
Theoretical calculations show that the strong boron-boron covalent bonds and metal-boron bonds play a crucial role for their high stability \cite{HEB2}.
So far, several single phase HEBs are reported \cite{HEB1}, including (Ti$_{0.2}$Zr$_{0.2}$Hf$_{0.2}$Nb$_{0.2}$Ta$_{0.2}$)B$_{2}$ , (Ti$_{0.2}$Zr$_{0.2}$Hf$_{0.2}$Mo$_{0.2}$Ta$_{0.2}$)B$_{2}$,
(Ti$_{0.2}$Zr$_{0.2}$Hf$_{0.2}$Nb$_{0.2}$
Mo$_{0.2}$)B$_{2}$, and (Ti$_{0.2}$Zr$_{0.2}$Hf$_{0.2}$Cr$_{0.2}$Ta$_{0.2}$)B$_{2}$.
These HEBs can be obtained in either bulk or thin-film form \cite{HEB3} by various techniques \cite{HEB4,HEB5,HEB6,HEB7,HEB8,HEB9}.
Intriguingly, the hardness and oxidation resistance of the HEBs are better than the average of the individual binary counterparts \cite{HEB1,HEB6,HEB10}, and their chemical tunability opens a wide space to explore the composition-property relationship in multicomponent diborides \cite{HEB11,HEB12}.

The presence of metal deficiency is well documented in some binary AlB$_{2}$-type transition metal diborides, which not only affects the lattice constants but also changes their properties.
Actually, this was firstly identified in prototype AlB$_{2}$ and hence its chemical formula is better expressed as Al$_{1-\delta}$B$_{2}$ with $\delta$ up to $\sim$0.1 \cite{AlB2-1,AlB2-2,AlB2-3}.
Later on, metal deficient Nb$_{1-\delta}$B$_{2}$ \cite{NbB2-1,NbB2-2,NbB2-3}, Ta$_{1-\delta}$B$_{2}$ \cite{NbB2-1} and (Mo$_{1-x}$$T$$_{x}$)$_{1-\delta}$B$_{2}$ ($T$ = Zr and Sc) \cite{MoZrB2,MoScB2} have also been synthesized and characterized.
Especially, $\delta$ reaches $\sim$0.5 for the Nb$_{1-\delta}$B$_{2}$ samples prepared under high pressure \cite{NbB2-1}.
On the contrary, TiB$_{2}$, ZrB$_{2}$ and HfB$_{2}$ are either stoichiometry compounds or have a rather small homogeneity range ($\delta$ $<$ 0.03) \cite{NbB2-3}.
In this context, a natural question to ask is whether metal deficiencies can exist in the HEB such as (Ti$_{0.2}$Zr$_{0.2}$Hf$_{0.2}$Nb$_{0.2}$Ta$_{0.2}$)B$_{2}$.
To our knowledge, however, no such investigation has been carried out to date.

Here we present the study of the metal deficient (Ti$_{0.2}$Zr$_{0.2}$Hf$_{0.2}$Nb$_{0.2}$Ta$_{0.2}$)$_{1-\delta}$B$_{2}$ HEBs.
Structural analysis indicates the formation of a pure and homogeneous AlB$_{2}$-type phase for 0.03 $\leq$ $\delta$ $\leq$ 0.18.
With increasing $\delta$, the $a$-axis shrinks continuously while the $c$-axis first increases then decreases.
The Vickers microhardnesses of these HEBs are measured and their $\delta$ dependence is examined.
Also, the phase stability is discussed in relation to the presence of metal deficiency.
\begin{figure*}
\includegraphics*[width=17.4cm]{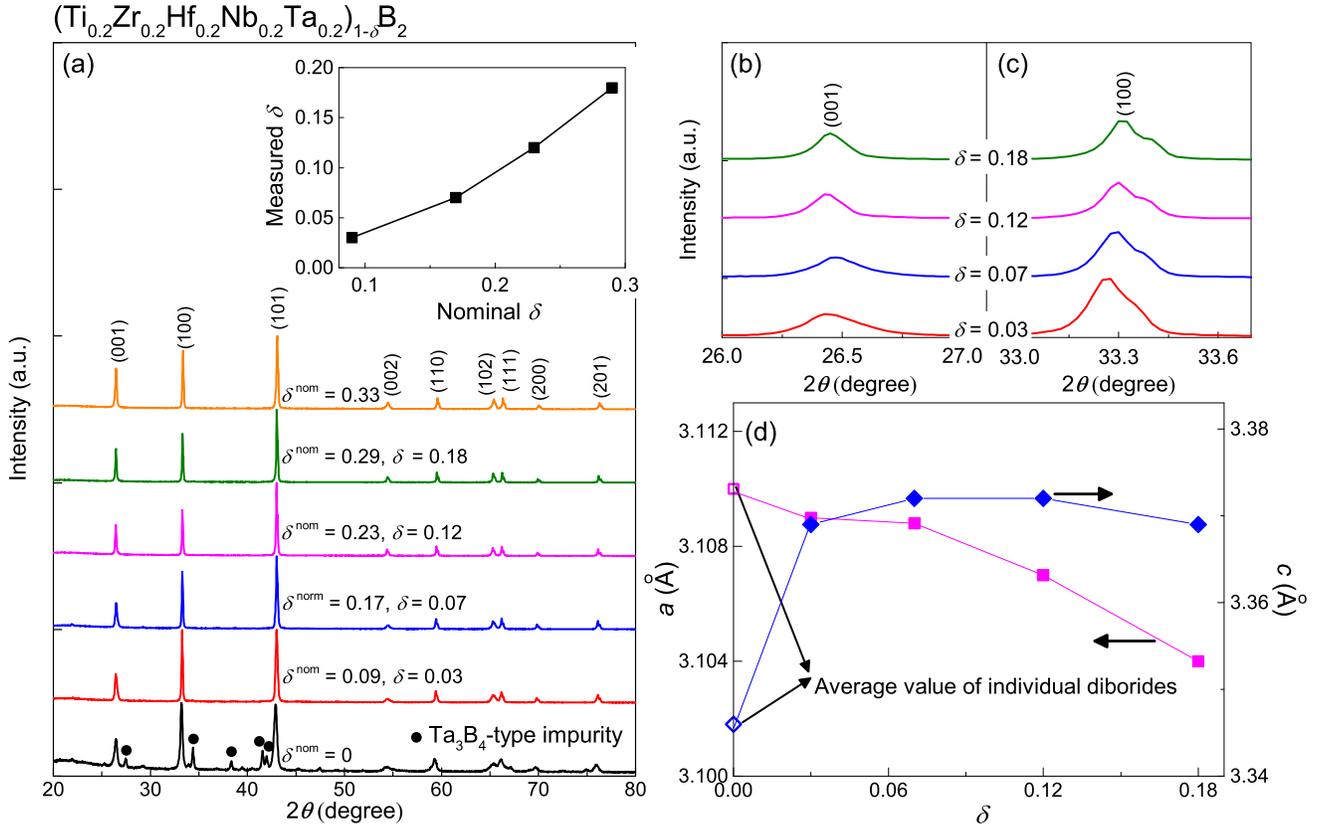}
\caption{
(a)  XRD patterns for the series of (Ti$_{0.2}$Zr$_{0.2}$Hf$_{0.2}$Nb$_{0.2}$Ta$_{0.2}$)$_{1-\delta}$B$_{2}$ HEBs. The nominal and measured $\delta$ values are labeled whenever possible. The impurity peaks for $\delta^{\rm nom}$ = 0 are labeled and the diffraction peaks for $\delta^{\rm nom}$ = 0.33 are indexed based on the AlB$_{2}$-type structure. The inset shows the measured $\delta$ plotted as a function of nominal $\delta$.
(b, c) Zooms of the (001) and (100) diffraction peaks for 0.03 $\leq$ $\delta$ $\leq$ 0.18, respectively. (d) Hexagonal lattice parameters plotted as a function of $\delta$. The error bars are within the size of symbols.
}
\label{fig1}
\end{figure*}

\section{2. Experimental section}
The (Ti$_{0.2}$Zr$_{0.2}$Hf$_{0.2}$Nb$_{0.2}$Ta$_{0.2}$)$_{1-\delta}$B$_{2}$ HEBs with nominal 0 $\leq$ $\delta$ $\leq$ 0.33 were prepared by the arc-melting method. High purity transition metal (99.99\%) and boron (99.99\%) elements with a total mass of $\sim$200 mg were weighed according to the stoichiometric ratio, mixed thoroughly and pressed into pellets in an argon-filled glovebox. The pellets were then melted in an arc furnace under high-purity argon atmosphere (99.999\%) with a current of 80 A. The melts were flipped and then remelted for several times to ensure homogeneity, followed by rapid cooling on a water-chilled copper plate. The as-cast HEBs are ball shaped with a radius of $\sim$4 mm and a volume of $\sim$33.5 mm$^{3}$.
The crystal structure of resulting samples was checked by powder x-ray diffraction (XRD) using a Bruker D8 Advance x-ray diffractometer with Cu-K$\alpha$ radiation at room temperature.
Neither external nor internal standard was used in the XRD measurements and the lattice parameters are determined by the least-squares fitting using the powderX software. The elemental composition was determined by inductively coupled plasma optical emission spectroscopy (ICP-OES) since this method enables an accurate determination of the boron content \cite{ICP-AES}. The details of ICP-OES measurements and raw elemental content data can be found in the Supplementary Material. The morphology and elemental distribution were investigated by a Zeiss Supratm 55 Schottky field emission scanning electron microscope (SEM) equipped with an energy dispersive x-ray (EDX) spectrometer.
All images were taken in the secondary electron mode.
The microstructure was examined in an FEI Tecnai G2 F20 S-TWIN transmission electron microscope (TEM) operated under an accelerating voltage of 200 kV.
The Vickers micro-hardness $H_{\rm V}$ was measured using an automatic microhardness testing system with applied loads $P$ varying from 0.49 N to 9.8 N, and $H_{\rm V}$ is given as
\begin{equation}
H_{\rm V} = 1854.4P/r^{2},
\end{equation}
where $r$ is the mean length of the two indention diagonals. For each $P$, the data are averaged from three independent measurements and the dwelling time for each measurement is 10 s.

\begin{figure*}
\includegraphics*[width=17.6cm]{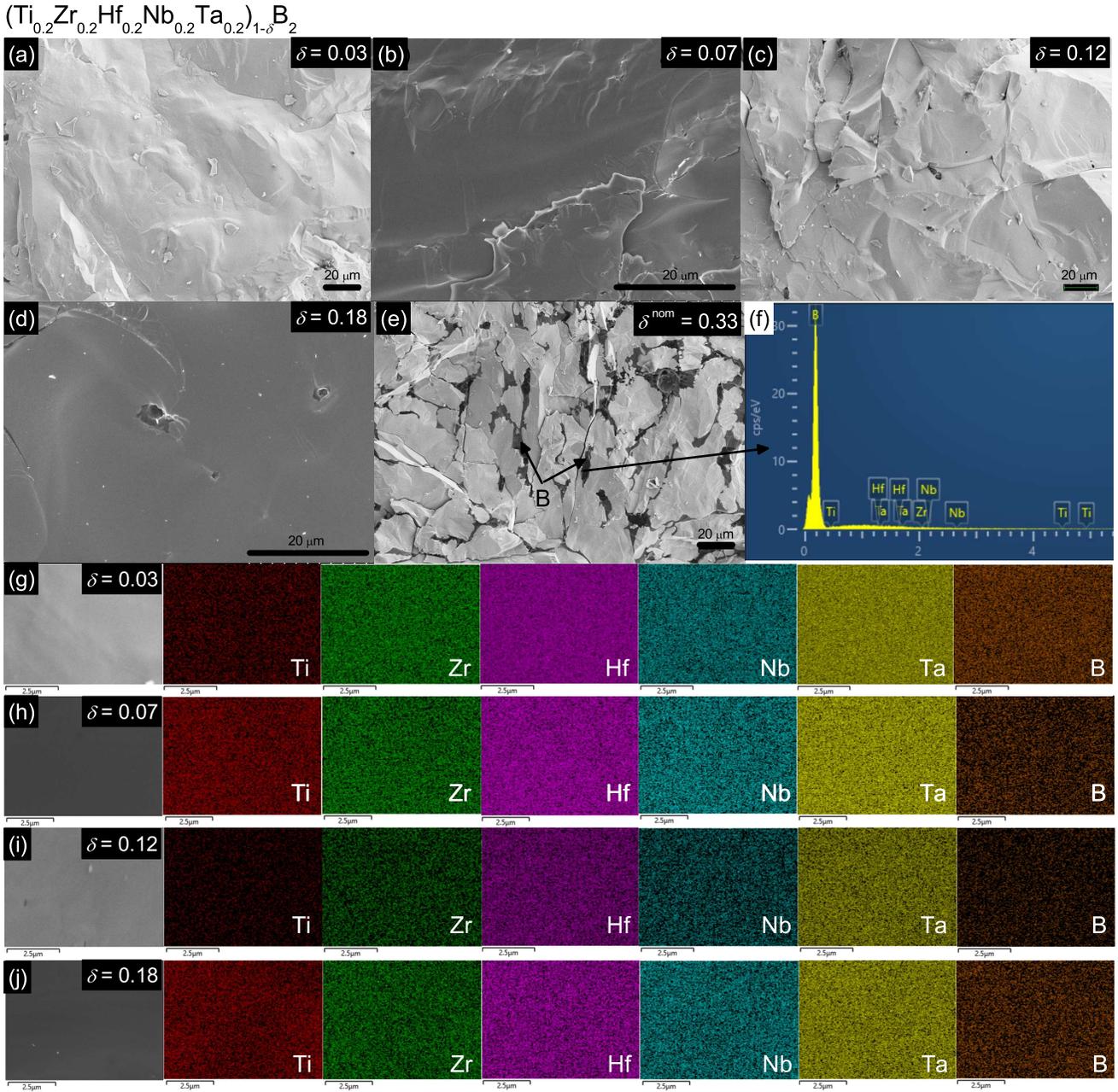}
\caption{
(a-e) SEM images on a scale bar of 20 $\mu$m for the (Ti$_{0.2}$Zr$_{0.2}$Hf$_{0.2}$Nb$_{0.2}$Ta$_{0.2}$)$_{1-\delta}$B$_{2}$ HEBs with $\delta$ = 0.03, 0.07, 0.12, 0.18 and $\delta^{\rm nom}$ = 0.33.
In panel (e), the arrows mark the boron impurity, whose EDX spectra is shown in panel (f).
(g-j) SEM image on a scale bar of 2.5 $\mu$m and corresponding EDX elemental maps for the HEBs with 0.03 $\leq$ $\delta$ $\leq$ 0.18.
}
\label{fig1}
\end{figure*}

\section{3. Results and Discussion}
The XRD patterns for the series of (Ti$_{0.2}$Zr$_{0.2}$Hf$_{0.2}$Nb$_{0.2}$Ta$_{0.2}$)$_{1-\delta}$B$_{2}$ HEBs are displayed in Fig. 1(a) with increasing nominal metal deficiency $\delta^{\rm nom}$ from bottom to top.
For $\delta^{\rm nom}$ = 0, the HEB has a major AlB$_{2}$-type phase but contains the orthorhombic Ta$_{3}$B$_{4}$-type impurity phase, whose amount is significantly higher than that obtained by the spark plasma sintering \cite{HEB1}. Since this impurity phase has a boron-to-metal ratio less than 2:1, it is reasonable to speculate that its formation is due to the lack of boron source.
Indeed, as metal deficiencies are introduced, a single AlB$_{2}$-type phase is retained up to $\delta^{\rm nom}$ = 0.33, as evidenced by indexing the XRD peaks based on the $P$6/$mmm$ space group.
Nevertheless, as shown in the inset of Fig. 1(a) and Table I, the actual $\delta$ values measured by the ICP-OES method are considerably smaller than the nominal ones, indicating a significant sublimation of boron during the arc melting process. Specifically, with increasing $\delta^{\rm nom}$ from 0.09 to 0.29, the actual $\delta$ only increases from 0.03 to 0.18. In passing, the $\delta$ parameter is not determined for the HEB with $\delta^{\rm nom}$ = 0.33 since this sample contains boron impurity (see below).
\begin{figure*}
\includegraphics*[width=17.6cm]{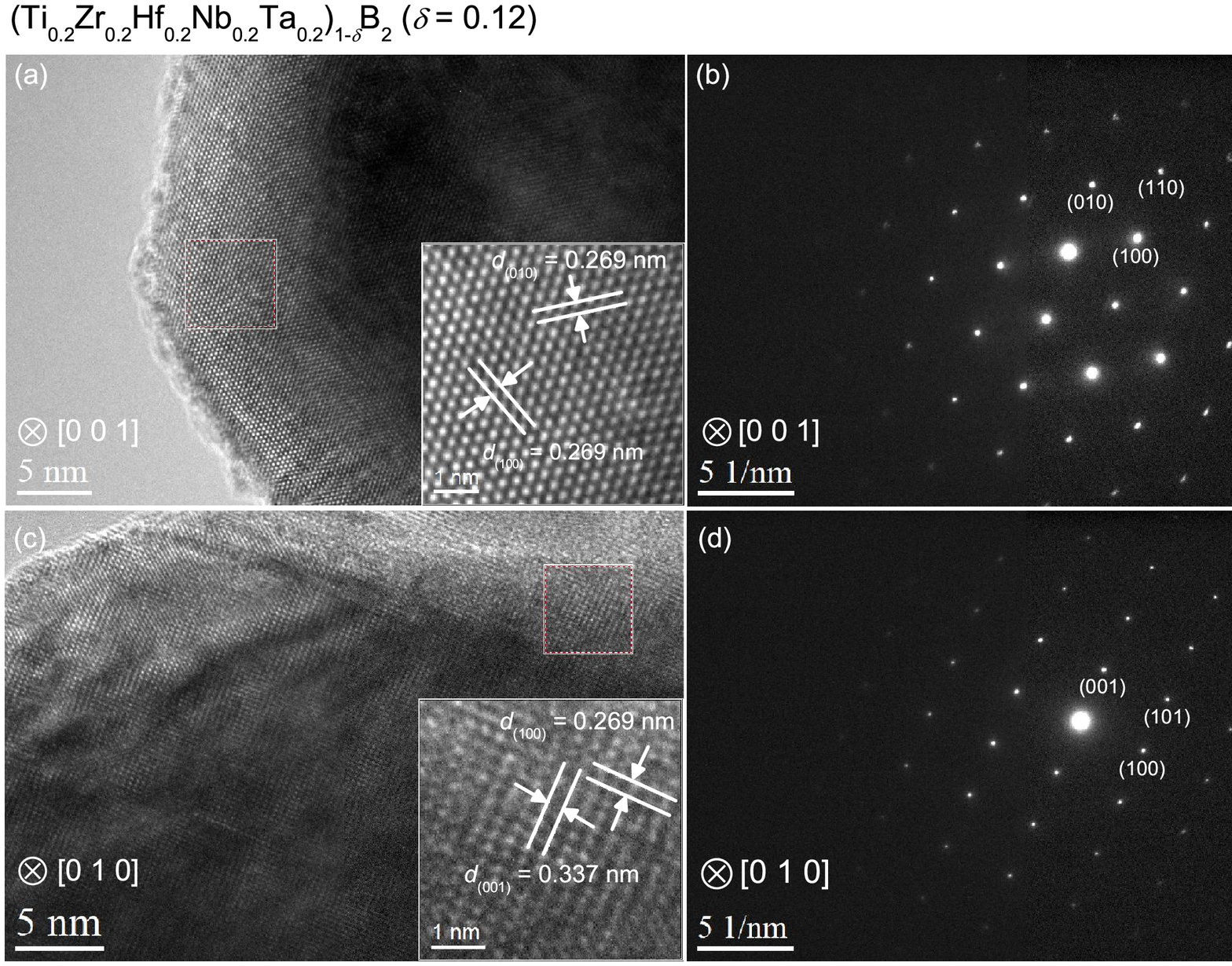}
\caption{
(a-b) TEM image and corresponding SAED pattern for the (Ti$_{0.2}$Zr$_{0.2}$Hf$_{0.2}$Nb$_{0.2}$Ta$_{0.2}$)$_{1-\delta}$B$_{2}$ HEB with $\delta$ = 0.12 taken along the [0 0 1] zone axis.
In panel (a), the inset shows a zoom of the TEM image and the lattice spacings are indicated.
(c-d) TEM image and corresponding SAED pattern for this HEB taken along the [0 1 0] zone axis.
In panel (c), the inset shows a zoom of the TEM image and the lattice spacings are indicated.
}
\label{fig1}
\end{figure*}

Figures 1(b) and (c) show the zooms of (001) and (100) peaks for HEBs with 0.03 $\leq$ $\delta$ $\leq$ 0.18.
As the increase of $\delta$, the (001) peak shifts nonmonotonically while the (100) peak moves steadily towards higher 2$\theta$ values (the bulging peak at 2$\theta$ $\approx$ 33.4$^{\circ}$ is the diffraction contribution from Cu K$\alpha$2 radiation).
This points to a nonmonotonic variation of the $c$-axis and a monotonic shrinkage of the $a$-axis.
These trends are confirmed by the $\delta$ dependence of the hexagonal lattice parameters shown in Fig. 1(d).
For $\delta$ = 0.03, we obtain $a$ = 3.111(1) {\AA} and $c$ = 3.359(1) {\AA}.
With increasing $\delta$, the $a$-axis length decreases continuously to 3.102(1) {\AA} at $\delta$ = 0.18 while the $c$-axis length shows a maximum of 3.372(1) {\AA} at $\delta$ = 0.07 and 0.12.
As a consequence, the unit-cell volume reaches a maximum of 28.22 {\AA}$^{3}$ at $\delta$ = 0.07. It is worth noting that the presence of metal vacancies reduces both the average radius and positive charge density of the transition metal layers. The former tends to shorten the $a$- and $c$-axis, while the latter tends to weaken the electrostatic force between the transition metal and boron layers hence increases the $c$-axis.
The interplay between these two effects qualitatively explains the variation of lattice parameters observed experimentally. Interestingly, the lattice parameters of (Ti$_{0.2}$Zr$_{0.2}$Hf$_{0.2}$Nb$_{0.2}$Ta$_{0.2}$)B$_{2}$ ($\delta$ = 0) obtained by averaging the individual metal diborides ($a$ = 3.110(1) {\AA} and $c$ = 3.346(1) {\AA}) \cite{HEB1} follow well this trend. In contrast, the experimental values for $\delta$ = 0 (see Table I) are more close to the (Ti$_{0.2}$Zr$_{0.2}$Hf$_{0.2}$Nb$_{0.2}$Ta$_{0.2}$)$_{1-\delta}$B$_{2}$ HEBs with $\delta$ $\sim$ 0.07-0.12. It is thus tempting to speculate that metal deficiencies are already present in these HEB samples.

\begin{figure*}
\includegraphics*[width=17.6cm]{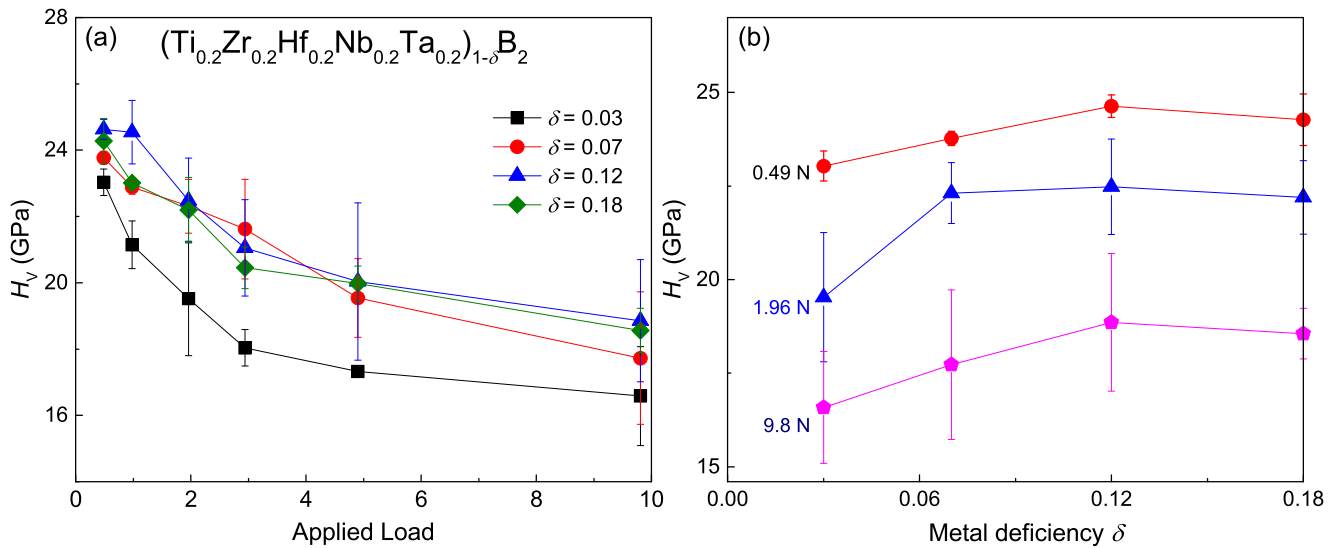}
\caption{
(a) Dependence of Vickers hardness on the applied load for the series of (Ti$_{0.2}$Zr$_{0.2}$Hf$_{0.2}$Nb$_{0.2}$Ta$_{0.2}$)$_{1-\delta}$B$_{2}$ HEBs.
(b) Dependence of Vickers hardness on the metal deficiency $\delta$ for applied loads of 0.49 N, 1.96 N and 9.8 N.
}
\label{fig1}
\end{figure*}
Figures 2(a-e) show the SEM images with a scale bar of 20 $\mu$m for the (Ti$_{0.2}$Zr$_{0.2}$Hf$_{0.2}$Nb$_{0.2}$Ta$_{0.2}$)$_{1-\delta}$B$_{2}$ samples.
For $\delta$ $\leq$ 0.18, the samples are free from pores and secondary phases.
Only at $\delta^{\rm nom}$ = 0.33, small darker regions are observed in between the large lighter ones and confirmed to be elemental boron by the EDX measurements [Fig. 2(f)].
Since no corresponding diffraction peak is visible in the XRD pattern, it is most likely that this excess boron exists in the amorphous state, in contrast to the case in metal-deficient (Mo$_{1-x}$Sc$_{x}$)$_{1-\delta}$B$_{2}$ \cite{MoScB2}.
Note that the atoms in amorphous boron are randomly arranged and scatter the x-ray light in random directions, which gives a broad hump and no sharp diffraction peaks in the XRD pattern. Hence the amount of amorphous boron cannot be retrieved by the XRD result.
The SEM images and EDX elemental maps with a smaller scale bar of 2.5 $\mu$m for the HEBs with $\delta$ $\leq$ 0.18 are displayed in Figs. 2(g-j). These results clearly indicate that the Ti, Zr, Hf, Nb, Ta and B elements are distributed uniformly in both samples.
Furthermore, the ratios of Ti:Zr:Hf:Nb:Ta determined by ICP-OES analysis are in line with the nominal ones within the experimental error (see Table I), indicating that the equimolar metal composition is unaffected by the presence of metal deficiency.
One may note that, compared with the HEB with $\delta$ = 0.03, the Hf content of the HEB with $\delta$ = 0.12 is 0.01 higher while its Nb content is 0.01 lower. If the change of lattice parameters was caused by the variation of Ti:Zr:Hf:Nb:Ta ratios, one would expect that the lattice parameters of the latter are larger than those of the former since the lattice parameters of HfB$_{2}$ are larger than those of NbB$_{2}$ \cite{HEB1}. This is at odd with the experimental observation and hence it is more plausible that the change of lattice parameters is mainly due to the increase of metal deficiencies. In passing, from the electron back-scatter diffraction orientation maps (see Supplementary Material Fig. S1), the average grain sizes are estimated to be 235.9$\pm$120.9 $\mu$m, 128.4$\pm$71.7 $\mu$m, 171.9$\pm$139 $\mu$m, and 52.7$\pm$35.4 $\mu$m for the HEBs with $\delta$ = 0.03, 0.07, 0.12 and 0.18, respectively.

The (Ti$_{0.2}$Zr$_{0.2}$Hf$_{0.2}$Nb$_{0.2}$Ta$_{0.2}$)$_{1-\delta}$B$_{2}$ sample with $\delta$ = 0.12 was also characterized by TEM measurements.
Figure 3(a) shows the high resolution TEM (HRTEM) image taken along the [0 0 1] zone axis.
Clear lattice fringes are observed, indicative of a highly crystalline nature.
On zooming-in [the inset of Fig. 3(a)], one can see that the nearest neighbour atoms form triangular lattices with an equal lattice spacing of 0.269 nm,
corresponding to (100) and (010) planes of the hexagonal lattice.
As can be seen from Fig. 3(b), the selected-area electron diffraction (SAED) exhibits a well-defined spot pattern, and the spots near the center are well indexed to the (100), (010) and (110) planes.
The HRTEM image and SAED pattern taken along the [0 1 0] zone axis are shown in Figs. 3(c) and (d), respectively.
In this case, the nearest neighbour atoms form rectangular lattices with two different spacings of 0.269 nm and 0.337 nm, which match well with the (100) and (001) planes.
In addition, the spots near the center of SAED pattern are indexed to (100), (101) and (001) planes.
These results, combined with the SEM ones, affirm that this HEB is homogeneous on both a macroscopic and microscopic scale.

\begin{table*}
\caption{Chemical compositions, lattice parameters, Vickers hardnesses, average grain sizes and relative densities for the (Ti$_{0.2}$Zr$_{0.2}$Hf$_{0.2}$Nb$_{0.2}$Ta$_{0.2}$)$_{1-\delta}$B$_{2}$ HEBs.}
\renewcommand\arraystretch{2.0}
\begin{tabular}{p{2.2cm}<{\centering}p{6.9cm}<{\centering}p{0.9cm}<{\centering}p{0.9cm}<{\centering}p{1.2cm}<{\centering}p{1.5cm}<{\centering}p{1.6cm}<{\centering}p{1.4cm}<{\centering}}
\\
\hline% In-table horizontal line
Nominal $\delta$ & Actual composition & $a$  ({\AA})	&  $c$   ({\AA})     & $V$ ({\AA}$^{3}$)  & $H_{\rm V}^{9.8 \rm N}$ (GPa) & Grain size ($\mu$m)& Relative density\\
\hline% In-table horizontal line
$\delta^{\rm nom}$ = 0 & $-$ & 	3.111	 & 3.359 & 28.15	& $-$ & $-$ & $-$ \\
$\delta^{\rm nom}$ = 0.09 &(Ti$_{0.22}$Zr$_{0.20}$Hf$_{0.18}$Nb$_{0.21}$Ta$_{0.19}$)$_{0.97}$B$_{2}$ ($\delta$ = 0.03) & 	3.109 	 & 3.369 & 28.20 	& 16.6$\pm$1.4& 235.9$\pm$120.9 & 90.2\%\\
$\delta^{\rm nom}$ = 0.17 &(Ti$_{0.21}$Zr$_{0.20}$Hf$_{0.20}$Nb$_{0.20}$Ta$_{0.19}$)$_{0.93}$B$_{2}$ ($\delta$ = 0.07)& 	3.109 	 & 3.372 & 28.22 & 17.7$\pm$2 &128.4$\pm$71.7 & 85.3\%\\
$\delta^{\rm nom}$ = 0.23 &(Ti$_{0.22}$Zr$_{0.20}$Hf$_{0.19}$Nb$_{0.20}$Ta$_{0.19}$)$_{0.88}$B$_{2}$ ($\delta$ = 0.12)& 	3.107	 & 3.372 & 28.19  &18.9$\pm$1.8	&171.9$\pm$139 & 88.4\%\\
$\delta^{\rm nom}$ = 0.29 &(Ti$_{0.22}$Zr$_{0.19}$Hf$_{0.19}$Nb$_{0.20}$Ta$_{0.20}$)$_{0.82}$B$_{2}$ ($\delta$ = 0.18)&   3.104    & 3.369 & 28.12 &18.6$\pm$0.7 &52.7$\pm$35.4 & 92.2\%\\
$\delta^{\rm nom}$ = 0.33 &$-$ &   3.102    & 3.367 & 28.05 & $-$ & $-$ & $-$	\\
\multicolumn{2}{c}{Average values of individual metal diborides}   & 	3.110 	 & 3.346 & $-$	& 18.4$\pm$1.1 & $-$ & $-$\\
\multicolumn{2}{c}{(Ti$_{0.2}$Zr$_{0.2}$Hf$_{0.2}$Nb$_{0.2}$Ta$_{0.2}$)B$_{2}$  \cite{HEB1}}   & 	3.101 	 & 3.361	& $-$& 17.5$\pm$1.2& $-$& 92.4\%\\
\multicolumn{2}{c}{(Ti$_{0.2}$Zr$_{0.2}$Hf$_{0.2}$Nb$_{0.2}$Ta$_{0.2}$)B$_{2}$  \cite{HEB10}}   & 	3.105 	 & 3.378 	& $-$& 21.0$\pm$0.2& 4.9$\pm$2.4& 99.5\%\\
\hline% In-table horizontal line
\hline % Bottom horizontal line
\end{tabular}
\label{Table3}
\end{table*}
Figure 4(a) shows the applied load $P$ dependence of $H_{\rm V}$ for the (Ti$_{0.2}$Zr$_{0.2}$Hf$_{0.2}$Nb$_{0.2}$Ta$_{0.2}$)$_{1-\delta}$B$_{2}$ samples with 0.03 $\leq$ $\delta$ $\leq$ 0.18.
For all $\delta$ values, $H_{\rm V}$ shows a common asymptotic behavior with increasing $P$ from 0.49 N to 9.8 N, which can be understood by the penetration depth effect \cite{MoRe2C}.
At low $P$, only the surface layer is affected by the indenter. However, at high $P$, the inner surface plays a dominant role and finally $H_{\rm V}$ becomes saturated.
Compared with $\delta$ = 0.03, the $H_{\rm V}$-$P$ curves appear to shift up for higher $\delta$ values.
To examine the effect of metal deficiencies on the hardness, we plot the same set of data as a function of $\delta$ in Fig. 4(b).
For almost all $P$ values, a shallow maximum is observed at $\delta$ = 0.12.
At 9.8 N, the corresponding $H_{\rm V}$ achieves 18.9 GPa , which is 14\% larger than that (16.6 GPa) at $\delta$ = 0.03.
Nevertheless, this trend should be taken with caution since the hardness can also be affected by extrinsic factors such as grain size, relative density and inclusions.
Indeed, it is noted that the $H_{\rm V}$ values of our HEB samples are smaller than some of those reported for the (Ti$_{0.2}$Zr$_{0.2}$Hf$_{0.2}$Nb$_{0.2}$Ta$_{0.2}$)B$_{2}$ HEB \cite{HEB6,HEB10}.
This is understandable since the latter preparation involves solid-state sintering under pressure, which results in smaller grain sizes and higher relative densities than those in the present case (see Table I).
Hence future studies are needed to draw a more definitive conclusion about the correlation between $H_{\rm V}$ and the parameters including the unit-cell volume and metal deficiency in the (Ti$_{0.2}$Zr$_{0.2}$Hf$_{0.2}$Nb$_{0.2}$Ta$_{0.2}$)$_{1-\delta}$B$_{2}$ HEBs.

We now discuss the role of metal deficiency in the phase stability of (Ti$_{0.2}$Zr$_{0.2}$Hf$_{0.2}$Nb$_{0.2}$Ta$_{0.2}$)$_{1-\delta}$B$_{2}$ HEBs.
In general, the stability of high-entropy systems is closely related to the parameters including atomic size difference and mixing entropy ($\Delta$$S_{\rm mix}$) \cite{HEB1,stability1,stability2},
and a single phase is favored at a small atomic size difference and a large $\Delta$$S_{\rm mix}$.
For AlB$_{2}$-type HEBs, both quantities are essentially determined by the transition metal sublattice.
The introduction of metal vacancies has practically no effect on the atomic size difference but significantly affects the mixing entropy,
which can be expressed in terms of $\delta$ as
\begin{equation}
\Delta S^{\rm mix}/R = 1.61(1-\delta)-(1-\delta)\rm ln(1-\delta)-\delta \rm ln \delta
\end{equation}
and $R$ is gas constant.
The calculated $\Delta$$S_{\rm mix}$ equals to 1.61$R$ for $\delta$ = 0, but rises to 1.75-1.79$R$ for $\delta$ in the range of 0.03 to 0.18.
It is thus reasonable to speculate that the increased $\Delta$$S_{\rm mix}$ is mainly responsible for stabilizing the single phase metal deficient (Ti$_{0.2}$Zr$_{0.2}$Hf$_{0.2}$Nb$_{0.2}$Ta$_{0.2}$)$_{1-\delta}$B$_{2}$ HEBs.
Note that this mechanism is irrelevant to the kinds of transition metal elements, and hence is expected to be general for AlB$_{2}$-type HEBs, the verification of which is certainly of interest for future studies.
\section{4. Conclusion}
In summary, we have successfully obtained phase pure AlB$_{2}$-type (Ti$_{0.2}$Zr$_{0.2}$Hf$_{0.2}$Nb$_{0.2}$Ta$_{0.2}$)$_{1-\delta}$B$_{2}$ HEBs with metal deficiencies $\delta$ ranging from 0.03 to 0.18.
The increase of $\delta$ leads to a continuous decrease in $a$-axis and a nonmonotonic variation of the $c$-axis.
As a consequence, the unit-cell volume displays a maximum at $\delta$ = 0.07.
These metal-deficient HEBs have high Vickers hardness of 16.6-18.9 GPa at 9.8 N and are likely stabilized by the increased mixing entropy due to presence of metal deficiency.
Our study not only presents the first series of metal deficient AlB$_{2}$-type HEBs, but also calls for further studies to explore similar multicomponent diborides.

\section*{ACKNOWLEGEMENT}
We thank the foundation of Westlake University for financial support and the Service Center for Physical Sciences for technical assistance in SEM measurements.


\begin{thebibliography}{00}

\bibitem{HEB1}
J. Gild, Y. Y. Zhang, Y. Harrington, S. C. Jiang, T. Hu, M. C. Quinn, W. M. Mellor, N. X. Zhou, K. Vecchio, and J. Luo,
Sci. Rep. {\bf 6}, 37946 (2016).

\bibitem{HEB2}
Y. P. Wang, G. Y. Gan, W. Wang, Y. Yang, and B. Y. Tang,
Phys. Status Solidi B {\bf 255}, 1800011 (2018).

\bibitem{HEB3}
P. H. Mayrhofer, A. Kirnbauer, Ph. Ertelthaler, and C. M. Koller,
Scr. Mater. {\bf 149}, 93 (2018).

\bibitem{HEB4}
G. Tallarita, R. Licheri, S. Garroni, T. Orru, and G. Cao,
Scr. Mater. {\bf 158}, 100 (2019).

\bibitem{HEB5}
D. Liu, T. Q. Wen, B. L. Ye, and Y. H. Chu,
Scr. Mater. {\bf 167}, 110 (2019).

\bibitem{HEB6}
J. F. Gu, S. K. Sun, H. Wang, S. Y. Yu, J. Y. Zhang, W. M. Wang, and Z. Y. Fu,
Sci. China Mater. {\bf 62}, 1898 (2019).

\bibitem{HEB7}
S. Failla, P. Galizia, S. Fu, S. Grasso, and D. Sciti,
J. Eur. Ceram. Soc. {\bf 40}, 588 (2020).

\bibitem{HEB8}
J. Gild, A. Wright, K. Quiambao-Tmoko, M. Qin, J. A. Tomko, M. S. bin Hoque, J. L. Braun, B. Bloomfield, D. Martinez, T. Harrington, K. Vecchio, P. E. Hopkins, and J. Luo,
Ceram. Int. {\bf 46}, 6906 (2020).

\bibitem{HEB9}
Y. Zhang, S. K. Sun, W. Zhang, Y. You, W. M. Guo, Z. W. Chen, J. H. Yuan, and H. T. Lin,
Ceram. Int. {\bf 46}, 14299 (2020).

\bibitem{HEB10}
L. Feng, F. Monteverde, W. G. Fahrenholtz, and G. E. Hilmas,
Scr. Mater. {\bf 199}, 113855 (2021).

\bibitem{HEB11}
F. Monteverde, F. Saraga, M. Gabodardi, and J. R. Plaisier,
J. Eur. Ceram. Soc. {\bf 41}, 6255 (2021).

\bibitem{HEB12}
Z. Zhang, S. Z. Zhu, Y. B. Liu, L. Lin, and Z. Ma,
J. Eur. Ceram. Soc. {\bf 42}, 3685 (2022).

\bibitem{AlB2-1}
V. I. Matkovich, J. Economy, and R. F. Giese, Jr.,
J. Am. Chem. Soc. {\bf 84}, 2337 (1964).

\bibitem{AlB2-2}
I. Loa, K. Kunc, K. Syassen, and P. Bouvier,
Phys. Rev. B {\bf 66}, 134101 (2002).

\bibitem{AlB2-3}
U. Burkhardt, V. Gurin, F. Haarmann, H. Borrmann, W. Schnelle, A. Yaresko, and Y. Gurin,
J. Solid State Chem. {\bf 177}, 389 (2004).

\bibitem{NbB2-1}
Y. Yamamoto, C. Takao, T. Masui, M. Izumi, and S. Tajima,
Physica C {\bf 383}, 197 (2002).

\bibitem{NbB2-2}
I. R. Shein and A. L. Ivanovskii,
Physica C {\bf 468}, 2224 (2008).

\bibitem{NbB2-3}
C. A. Nunes, D. Kaczorowski, P. Rogl, M. R. Baldissera, P. A. Suzuki, G. C. Coelho, A. Grytsiv, G. Andre, F. Bouree, and S. Okada,
Acta Mater. {\bf 53}, 3679 (2005).

\bibitem{ICP-AES}
R. Hahn, V. Moraes, A. Limbeck, P. Polcik, P. H. Mayrhofer, H. Euchner,
Acta Mater. {\bf 174}, 398 (2019).

\bibitem{MoZrB2}
L. E. Muzzy, M. Avdeev, G. Lawes, M. K. Haas, H. W. Zandbergen, A. P. Ramirez, J. D. Jorgensen, and R. J. Cava,
Physica C {\bf 382}, 153 (2002).

\bibitem{MoRe2C}
Y. W. Cui, J. F. Wu, B. Liu, Q. Q. Zhu, G. R. Xiao, S. Q. Wu, G. H. Cao, and Z. Ren,
J. Alloys Compd. {\bf 856}, 157314 (2021).

\bibitem{MoScB2}
W. Z. Yang, G. R. Xiao, Q. Q. Zhu, Y. W. Cui, S. J. Song G. H. Cao, and Z. Ren,
Ceram. Int. {\bf 48}, 19971 (2022).

\bibitem{stability1}
Y. Zhang, T. T. Zuo, Z. Tang, M. C. Gao, K. A. Dahmen, P. K. Liaw, and Z. P. Lu,
Prog. Mater. Sci. {\bf 61}, 1 (2014).

\bibitem{stability2}
M.-H. Tsai and J.-W. Yeh,
Mater. Res. Lett. {\bf 2}, 107 (2014).

\bibitem{ZrB2-x}
D. Kalish, E. V. Clougherty, and K. Kreder,
J. Am. Ceram. Soc. {\bf 52}, 30 (1969).

\bibitem{hardness1}
X. L. Gu, C. Liu, H. Guo, K. Zhang, and C. F. Chen,
Acta Mater. {\bf 207}, 116685 (2021).


\end{thebibliography}
\end{document}